# Towards Enabling Architectural Refactorings through Source Code Annotations

Holger Krahn, Bernhard Rumpe
Technische Universität Braunschweig
Institut für Software System Engineering
38106 Braunschweig, Germany

**Abstract:** It is well known that software needs to change to meet new requirements. The synchronization of software architecture models and implementation is of high importance to keep the architecture documents useful and the software evolution process manageable. In this paper we achieve this synchronization by a two-step process. First, we augment the source code with architectural information. Second, this "lightweight architectural model" can be checked more easily against the full architectural description. Based on this approach refactorings on either side (code or architecture) are detected automatically and conformance checks become possible.

## 1 Introduction

The conceptual architecture of a software system can be described by components and connectors. Furthermore, ports are used as well defined communication interfaces between components. Such a software architecture description is then used as a guideline for the structuring and implementation of a software system.

In current development projects, the architecture description is normally used to develop the initial version of the software. As the software evolves, the requirements change or the source code is refactored, it may happen that the architecture description is not updated accordingly. Experience reports from practicing architects argue that the software architecture is already out-dated in the moment it is published [SSWA96].

Therefore it is important to use refactorings on an architectural level in conformance with the existing source code. To keep both system descriptions (architecture and source code) synchronized, we describe a special form of annotating the source code such that the code remains coupled with its software architecture description. Ideally the source code will be refactored in synchronization with the architecture description by using the coupling via annotations. However, in a first step it is already useful that a developer can discover the affected parts of the source code automatically although the code still has to be changed manually.

Our hypothesis is that an architecture description is of more value, if it is explicitly present to all developers during the coding phase [Rum04a, Rum04b]. In agile development projects all developers need to understand every artifact and part of the software. In larger projects this goal can only be achieved, if the developers can rely on an abstract description of the whole system. Therefore condensed and accurate architecture descriptions are a key factor. As described earlier, this paper concentrates on an approach to keep architecture and code synchronized through source code annotations. These annotations are not meant to replace the architecture description but to keep it consistent with the code.

The rest of the paper is structured as follows. Section 2 explains the problem of architectural conformance and its relevance for architectural refactorings. In Section 3 the software architecture description used in this paper is shown. Section 4 describes briefly the annotations provided for a developer, whereas Section 5 explains how this annotations can be used to automatically check architectural conformance. Section 6 shows which kind of refactorings can be applied to our architectural description. Section 7 compares our approach to related work and Section 8 concludes this paper.

## 2 Architectural Conformance

The source code and its software architecture description are seperate artifacts, both produced and used in different activities of the development cycle. Thus both artifacts usually evolve independently and it is very doubtful that both stay consistent with each other without considerable effort. This effort is usually called checking for architectural conformance [Men00] and needs to be repeated when evolving one of the mentioned artifacts.

Architectural reviews make architectural conformance possible. In order to assist and simplify this process we propose to use an intermediate description of the architecture which is added to the source code in form of annotations. The intermediate description can be checked in an automatic fashion against the architectural description. This check can be applied repeatedly by a developer during coding. This approach is similar to writing unit tests and using a test framework to quickly check if the test cases still execute successfully after a code change.

However, the intermediate description is not necessarily consistent with the source code, as the annotations might be wrong or incomplete. This has to be checked during an architectural review phase. The proposed approach reduces the necessary effort, because only the relation between code and annotations has to be checked and not the relation between architecture description and code anymore. Furthermore, code and annotations are co-located in the source files which makes it more likely that both are updated together and that the programmer is well informed about the part of the system under change. Figure 1 illustrates our approach.

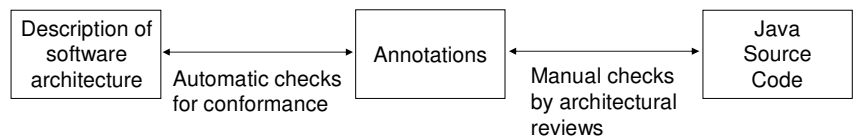

Figure 1: Architectural conformance in two steps

## 3 UML Composite Structure Diagrams

Software architectures can be described in various ways. We concentrate on Component-and-Connector (C&C) views [CBB+03] which can be used to model the structure of software systems. Often ports are added to a C&C view to describe well defined interaction points between components that are linked via connectors. C&C views are supported by most of the available architecture description languages; for a survey of the most common notations see [MT00].

In the new UML 2.0 standard [OMG] Composite Structure Diagrams (CSD) are introduced as a technique to model hierarchy, distribution and communication channels. In addition, roles can be associated with classes in order to model design patterns [GHJV96]. CSDs contain *Structured Classifiers* which are subclasses of the UML metaclass *Classifier* and allow classes to have parts and ports. A part has an associated role name and a type. The relationship between a part and its class can be understood as an aggregation. Ports are interaction points that isolate the classifier from its environment [RSRS99]. One of the advantages of UML CSDs is the possibility to model classes and instances of classes in a uniform way. CSDs allow us to describe the connections between their parts on a more detailed level than class diagrams but are more general than object diagrams which only describe exemplary instances. Figure 2 contains such a CSD.

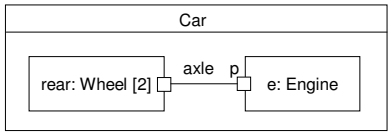

Figure 2: Example CSD [OMG]

## 4 Annotating Java Source Code

Java in its current version 5.0 [Jav] extends the pure programming language by adding annotations. This approach substitutes older annotation forms that used comments in the source code (e.g. [WR03]). The developer can annotate packages, classes, interfaces, enums, fields, constructors, methods, method or constructor parameters, local variables and annotations themselves, but no other language elements. The annotation developer can restrict the annotation to certain elements of the list above. In addition three retention policies are possible. The option ranges from pure checking by the compiler and discarding the annotation afterwards, over storing the annotation in the class file but not making it available at runtime, to full availability at runtime and accessibility through the reflection mechanism.

In our approach, we supply the programmer with a set of predefined annotations to relate the source code to a software architecture description. As discussed above, the usage of the annotations is similar to writing unit tests during coding.

The `@Component` annotation assigns a class or interface to a component in the architecture description. Figure 3 shows the annotation definition in Java source code that is restricted to types (which are mainly classes and interfaces). The retention policy is set to source because there is no reason to retain the annotations in the object code or even at runtime. This retention policy is the same for all our annotations.

The value of the `@Component` annotation describes its name. For convenience, we use the annotation element called `value` allowing us to use the annotation in the form `@Component("ComponentName")` for a component called *ComponentName*.

```
import java.lang.annotation.*;

/** This is an annotation for a component in
 *  a software arch. description. */
public
  @Target({ElementType.TYPE})
  @Retention(RetentionPolicy.SOURCE)
  @interface Component {

    /** @return name of component */
    String[] value();
}
```

Figure 3: Java definition for the `@Component` annotation

To save space, we use an informal and condensed form to describe the following annotations that allow to model CSDs within Java source code. The following text contains the condensed information from Figure 3.

`@Component` applies to types
     `String value` : Name of the component

In some cases a class may belong to different components in different architectural views. Unfortunately Java doesn't allow us to use the same annotation type with a different value attributes for a single language element. Hence the value attribute is an array of strings. This use of arrays to store multiple values is the same for all following annotations but is omitted in the condensed form.

A part of a component is marked using the `@Part` annotation.

`@Part` applies to field declarations
     `String value` : Role of part in enclosing component

A port is a well-defined interaction point that a component exposes to its environment: (a) a groups of methods, (b) a supported interface, (c) a notification mechanism like in the observer pattern, et cetera.

`@Port` applies to methods, constructors and types
     `String value` : Name of the port it belongs to

During the initial specification of a software architecture, the parts of a component are often considered as static and it is often neglected how a component is created. This dynamic changes must be added during the implementation. A component usually does not initialize itself in an automatic fashion and in most cases it can be instantiated multiple times in a single system. Our annotations can be used to mark methods and constructors which add or remove parts from a component. Note that we decided to abstract away from a too detailed description: The annotations are not detailed enough to specify which instances are added or removed when the parts have a multiplicity different than one.

`@AddPart` applies to methods and constructors
    `String value` : Name of the part
    `String componentname` : Name of component

`@RemovePart` applies to methods and constructors
    `String value` : Name of the port
    `String componentname` : Name of component

In [SG96, p. 165] connectors are described as follows: "Connectors do not in general correspond individually to compilation units; they manifest themselves as table entries, instructions to a linker, dynamic data structures, system calls, initialization parameters [...] and the like." Especially light-weight connectors e.g. realized through method calls do not necessarily manifest in a direct structural connection. These are both difficult to discover during an architectural review and difficult to determine from the Java source code. Therefore, we decided that methods and constructors that setup or disconnect certain connectors should be annotated.

`@Connects` applies to methods and constructors
    `String leftcomponent` : Name of first component
    `String left` : Name of the part or port (opt)
    `String rightcomponent` : Name of second component
    `String right` : Name of the part or port (opt)
    `Enum[``LEFT",``RIGHT",``BIDIR"] type`: Connection type

In addition we provide two annotations that have the same form as the `@Connects` annotation. The `@Disconnects` annotation states that a method or constructor tears a connector down that connected two parts or ports.

The `@Connector` annotation states that a type, field declaration or local variable stores information concerning the connection between two other parts or ports. An example is a hashtable that manages a complex relationship between objects of two classes. The `@Connector` annotation is used for tagging the information store of existing connections whereas the `@Connects`/`@Disconnects` annotations are used to tag places where this connection changes. A possible implementation that conforms to the CSD from Figure 2 can be found in Figure 4.

```
  public @Component("Car") class Car {
    private @Part("rear") Wheel[] rear;

    private @Part("e") Engine e;

    public @AddPart({"rear", "e" })
      @Connects(left="rear",right="e.p",type=Arrow.LEFT)
      Car() { /*..*/ }
  }
```

Figure 4: Example implementation using annotations

## 5 A Sketch on Checking Architectural Conformance

The software architecture description of a system can be generated from the complete and correctly given set of annotations. However, for methodical reasons we don't recommend this procedure. One of the problems is that it is not necessarily wise to locate the information about where a class belongs only in the class itself. Then without tool support it becomes e.g. difficult to relocate the class into another component. Experience shows that it is better to maintain a separate software architecture description and check the actual architecture obtained from the annotations against it. Initially it is even useful to generate the code frames together with the respective annotations from appropriate architectural documents. However, when both artifacts exist, e.g. these checks can be applied:

- Completeness of annotations in the source code: For every model element in the CSD like components, parts and ports an annotations exist in the source code.

- Completeness of architecture description: For every annotation exists an equivalent element in the architecture description.

- Consistency of code and architecture description: Only connections listed in the architecture description exist in the source code.

Note that these checks only compare the annotations with the architecture description, but can be done automatically. The second step, to check the annotations with the source code itself, has to be done manually as explained earlier. This task can be supported by tools like SA4J [SA4] or JDepend [JD] but cannot be solved in an automatic fashion. Among others late binding, reflective language elements and the various manifestations of CSD modeling concepts in the code prevent us from a static analysis of the dependencies within arbitrary applications.

## 6 Refactorings on Architectural Level

An essential part of the refactoring concept is their triggering through "*bad smells*". A bad smell describes a constellation in a software system description that for some reason may cause trouble with respect to quality, maintainability, evolution or the like. Although Kent Beck and Martin Fowler coined the phrase that "no set of metrics rivals informed human intuition" [Fow99, p.75], we think that tool support is useful to detect potential smells and humans can easier judge on this pre-selection where potential flaws are located.

Additional information provided by the developer through annotations introduces some redundancy that can be used to detect inconsistencies. Good candidates are for example a group of classes associated with a component that is distributed among different packages or a single connector that has more than one connecting and disconnecting method (or none at all). These kinds of bad smells can be detected automatically by a tool and either (a) exposed to the developer or even (b) repaired automatically. In case of (a) the developer has to judge, if this bad smells are really flaws or constructs that are designed on purpose, while in (b) refactoring of source code affects the architectural design documents automatically.

The automatic architectural conformance check (cf. Section 5) can be applied to the system after each refactoring step to gain the confidence that the system is still conforming to its architectural description. This procedure is similar to and should also be supplemented by the execution of unit tests after each refactoring step.

Figure 5 describes a typical desktop application which consists of several parts that can be considered as top level components of a system. In order to extend it to a client-server application where some parts of the original system are executed on the server an architectural refactoring is planned. This is a rather simple example, but it demonstrate the idea of refactoring by composing basic steps to complex refactorings. We apply the following refactoring plan to this architecture. The results of the individual steps can also be found in Figure 6.

1.-4. Add four new ports to the system.

5.-7. Add new connectors to connect *Model* with *Query*.

8. Remove the obsolete connector from the system.

9.-10. Remove the obsolete ports from the system.

11. Split the system into two main components.

Now the resulting system has been split into two parts. At each step the developer can use the annotation lookup to find the elements in the source code that are related to elements of the architecture description and apply the necessary changes. To ensure that there is no behavioral difference between the new and the old system, it is necessary to ensure that the new connection and the old connection behave equally. This can e.g. be described through an invariant and can be tested through appropriate unit tests.

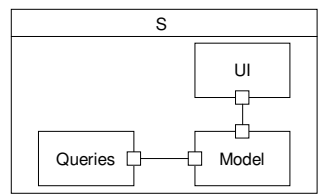

Figure 5: Software architecture description of the example system

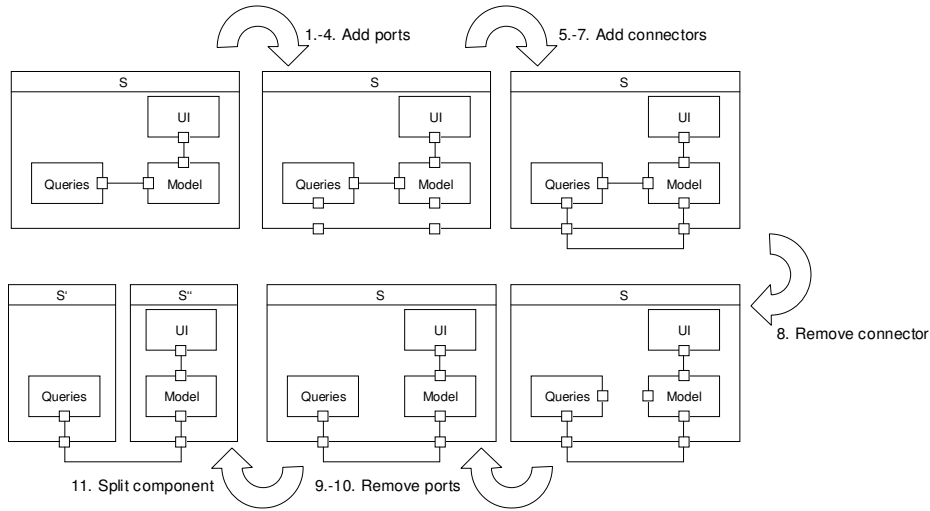

Figure 6: Complex refactoring on architectural level

After the refactoring a developer can e.g. change the communication technology of the connection from method calls to a TCP/IP-connection. This change is not reflected in the software architecture description because we modeled the logical communication on an abstract level only. The developer can list all methods that connect or disconnect a certain connection (as annotated by @Connects and @Disconnects) or list classes that represent the connection itself (as annotated by @Connector). This helps to localize the necessary changes which is especially helpful in larger systems.

## 7 Related Work

There are quite a number of approaches with similar goals and techniques. Here we concentrate on the most interesting ones. In [Men00] logical meta programming is used to describe the relationship between an implementation and different architectural views. The approach supports the user with automatic support for conformance checking of an architecture description and an implementation if the relationship is correctly specified. The approach differs from ours as it requires the use of an additional language, namely Prolog, to relate the source code with the architectural description. This leads to more expressivity as there are fewer restrictions on the elements that can be annotated, but requires more knowledge on technical details.

ArchJava [ACN02a, ACN02b] extends the Java programming language with first class concepts to describe connectors and ports. This approach allows the programmer to express the architecture directly in the program itself and thus makes conformance checking obsolete. However, using architectural diagrams directly as programming language has the disadvantage, that architectural planning and programming get intertwined, different views on a system cannot be provided and incomplete models are not allowed because this would directly lead to an incomplete software.

Gestalt [SSWA96] is an architecture description language that allows automatic conformance checks of an architecture description with source code. The approach is applied to the programming language C and checks interconnections by only using the include graph of source files.

In [MNS95] *software reflexion models* are introduced that relate source code elements to an architecture description. The system allows an automatic conformance check with information extracted from a call graph with a model that consists of modules and interconnections.

## 8 Summary and Outlook

The paper presents a lightweight approach to annotate Java source code using the newly Java 5.0 annotation mechanism to relate programming language constructs to a software architecture description. The annotations are coupled with the architecture description while the relationship between elements of source code and architecture can always be many to many. The main advantage is that this form of architecture description can easily be integrated in a code-centric agile development process, while also maintaining a separate architecture description.

The coupling of the software architecture description and the annotations is expressive enough to support the user when applying architectural refactorings. The approach allows predicting necessary code changes when applying a refactoring and checking consistence after restructuring the source code.

Our experiments show that the new Java annotation mechanism is generally helpful to relate code and architecture description. However, it would be even more helpful if we would be able to annotate more language elements and use the same annotation type more than once for a certain language element. We furthermore restricted ourselves to a subset of possible annotations which led to some restrictions when describing the architecture in the annotations. We are currently exploring if the proposed annotations are sufficient for practical use.

In the future we also want to explore if and how this kind of architecture description can be successfully used to automatically change larger parts of the existing source code. We want to statically analyze the existing source to allow a co-evolution of architecture description and source code planned on the architectural level. It is still an open question how reflective elements in the source code should be treated. Of course, this approach is not only interesting for software architecture but probably equally interesting for trac-ing requirements to code. We will look into this, when we have more experience with architecture and code links.

# References

[ACN02a] Jonathan Aldrich, Craig Chambers, and David Notkin. Architectural Reasoning in ArchJava. In *Proceedings of ECOOP '02*, pages 334–367. Springer-Verlag, 2002.

[ACN02b] Jonathan Aldrich, Craig Chambers, and David Notkin. ArchJava: connecting software architecture to implementation. In *Proceedings of ICSE '02*, pages 187–197, 2002.

[CBB$^+$03] Paul Clements, Felix Bachmnan, Len Bass, David Gralan, James Ivers, Reed Little, Robert Nord, and Judith Stafford. *Documenting Software Architectures - Views and Beyond*. Addison-Wesley, 2003.

[Fow99] Martin Fowler. *Refactoring: Improving the Design of Existing Programs*. Addison-Wesley, 1999.

[GHJV96] Erich Gamma, Richard Helm, Ralph Johnson, and John Vlissides. *Design Patterns*. Addison-Wesley, 1996.

[Jav] http://java.sun.com/j2se/1.5.0/.

[JD] JDepend. http://www.clarkware.com/software/JDepend.html.

[Men00] Kim Mens. *Automating architectural conformance checking by means of logic meta programming*. PhD thesis, Vrije Universiteit Brussel, 2000.

[MNS95] Gail C. Murphy, David Notkin, and Kevin Sullivan. Software reflexion models: bridging the gap between source and high-level models. In *Proc. of SIGSOFT '95*, pages 18–28. ACM, 1995.

[MT00] Nenad Medvidovic and Richard N. Taylor. A Classification and Comparison Framework for Software Architecture Description Languages. *IEEE Transactions on Software Engineering*, 2000.

[OMG] Object Management Group. *UML 2.0 Superstructure Specification*. http://www.uml.org/.

[RSRS99] Bernhard Rumpe, M. Schoenmakers, A. Radermacher, and Andy Schürr. UML + ROOM as a Standard ADL? In *Proceedings of ICECCS'99*, 1999.

[Rum04a] Bernhard Rumpe. *Agile Modellierung mit UML : Codegenerierung, Testfälle, Refactoring*. Xpert.press. Springer-Verlag, 2004.

[Rum04b] Bernhard Rumpe. *Modellierung mit UML : Sprache, Konzepte und Methodik*. Xpert.press. Springer-Verlag, 2004.

[SA4] Structural Analysis for Java. http://www.alphaworks.ibm.com/tech/sa4j.

[SG96] Mary Shaw and David Garlan. *Software Architecture - Perspectives on an Emerging Discipline*. Prentice Hall, 1996.

[SSWA96] R. W. Schwanke, V. A. Strack, and T. Werthmann-Auzinger. Industrial software architecture with Gestalt. In *Proceedings of IWSSD '96*, page 176. IEEE, 1996.

[WR03] Craig Walls and Norman Richards. *XDoclet in Action*. Manning, 2003.